\documentclass[12pt]{article}
\usepackage{amsmath}
\usepackage{amssymb}
\usepackage{epsfig}
\usepackage{amsthm}
\usepackage{stmaryrd}
\usepackage{bm}
\usepackage{hyperref}
\usepackage{eufrak}

\numberwithin{equation}{section}
\numberwithin{figure}{section}

\def\eq#1{(\ref{eq:#1})}
\def\lineup{\!\!\!\!\!\!\!\!&&}

\def\d{\partial}
\def\eps{\epsilon}

\def\fraction#1#2{ { \textstyle \frac{#1}{#2} }}
\def\half{\fraction{1}{2}}

\def\bare{bare}
\def\dressed{dressed}

\def\deg{\mathrm{deg}}
\def\Mb{\overline{M}}

\def\Mbb{\overline{\bf M}}
\def\M{{\bf M}}
\def\m{{\bf m}}
\def\Q{{\bf Q}}
\def\n{{\bm \eta}}
\def\b{{\bf b}}
\def\c{{\bf c}}
\def\g{{\bf g}}

\def\1st{{1\mathrm{st}}}

\begin{document}

\begin{titlepage}
\hfill LMU-ASC 82/13

\begin{center}

\vskip 1.5cm {\large \bf{Resolving Witten's Superstring Field Theory}}

\vskip 1.5cm

{\large Theodore Erler\footnote{tchovi@gmail.com}, Sebastian 
Konopka\footnote{sebastian.konopka@physik.uni-muenchen.de}, 
Ivo Sachs\footnote{ivo.sachs@physik.uni-muenchen.de}}

\vskip 1.0cm

{\it Arnold Sommerfeld Center, Ludwig-Maxmillians University,}\\
{\it Theresienstrasse 37, D-80333, Munich, Germany}
\vskip 1.5cm

{\bf Abstract}

\end{center}

We regulate Witten's open superstring field theory by replacing the 
picture-changing insertion at the midpoint with a contour integral of picture 
changing insertions over the half-string overlaps of the cubic vertex. The 
resulting product between string fields is non-associative, but we provide a 
solution to the $A_\infty$ relations defining all higher vertices. The result 
is an explicit covariant superstring field theory which by construction 
satisfies the classical BV master equation.

\noindent

\noindent
\medskip

\end{titlepage}

\tableofcontents

\section{Introduction}

For the bosonic string, the construction of covariant string field theories is 
more-or-less well understood. We know how to construct an action, 
quantize it, and prove that the vertices and propagators cover the the moduli 
space of Riemann surfaces relevant for computing amplitudes. For the 
superstring this kind of understanding is largely absent. A canonical 
formulation of open superstring field theory was provided by Berkovits 
\cite{BerkovitsI,BerkovitsII}, but it utilizes the ``large'' Hilbert 
space which obscures the relation to supermoduli space. Moreover, 
quantization of the Berkovits theory is not completely understood 
\cite{BerkPert,BerkPertBerk,Torii1,Torii2}. Motivated by this 
problem, we seek a different formulation of open superstring field theory 
satisfying three criteria:
\begin{description}
\item{(1)} The kinetic term is diagonal in mode number.
\item{(2)} Gauge invariance follows from the same algebraic structures which 
ensure gauge invariance in open bosonic string field theory.
\item{(3)} The vertices do not require integration over bosonic moduli.
\end{description}
We assume $(1)$ since we want the theory to have a simple 
propagator. We assume
$(2)$ since we want to be able to quantize the theory in a straightforward 
manner, following the work of Thorn \cite{Thorn}, Zwiebach 
\cite{ZwiebachClosed} and others for the bosonic string. Finally we 
assume $(3)$ for simplicity, but also because we would like to know 
whether open string field theory can describe closed string physics through its
quantum corrections. Once we add stubs to the open string vertices, the nature
of the minimal area problem changes and requires separate degrees of freedom 
for closed strings at the quantum level \cite{ZwOpCl}.

Condition $(1)$ rules out the modified cubic theory and its variants 
\cite{PTY,Russians,democratic,democraticgf,BerkSiegel,Kroyter_philosophy}, 
and $(2)$ rules out the Berkovits theory. This leaves the original proposal 
for open superstring field theory at picture $-1$, described by Witten 
\cite{Witten}. The problem is that this theory is singular 
and incomplete. A picture changing operator in the 
cubic term leads to a divergence in the four point amplitude which requires 
subtraction against a divergent quartic vertex \cite{Wendt}. Likely 
an infinite number of divergent higher vertices are needed to ensure 
gauge invariance, but have never been constructed.\footnote{There have been 
some attempts to fix the problems with Witten's theory by changing the 
nature of the midpoint insertions in the action. These include the modified
cubic theory \cite{PTY,Russians} and the theory described in \cite{Olaf}.}

In this paper we would like to complete the construction of Witten's open
superstring field theory in the NS sector. We achieve this by resolving the 
singularity in the cubic vertex by spreading the picture changing insertion 
away from the midpoint. As a result the product is non-associative. But we 
know how to formulate a gauge invariant action with a non-associative product 
\cite{Zwiebach}. The action takes the form
\begin{equation}S=\frac{1}{2}\omega(\Psi,Q\Psi) 
+\frac{1}{3}\omega(\Psi,M_2(\Psi,\Psi)) 
+\frac{1}{4}\omega(\Psi,M_3(\Psi,\Psi,\Psi))+  ...\ ,
\label{eq:action}\end{equation}
where $\omega$ is the symplectic bilinear form and $Q,M_2,M_3,...$ are 
multi-string products which satisfy the relations of an $A_\infty$ algebra. 
The fact that one can in principle construct a regularization of Witten's 
theory along these lines is well-known. The new ingredient we provide is an 
exact solution of the $A_\infty$ relations, giving an explicit and computable 
definition of the vertices to all orders.

The resulting theory is quite simple. However, its explicit form depends on a 
choice of BPZ even charge of the picture changing operator
\begin{equation}X = \oint \frac{dz}{2\pi i}f(z) X(z),\end{equation}
which tells us how to spread the picture changing insertion in the cubic 
vertex away from the midpoint. As far as we know, there is no canonical 
way to make this choice. This suggests the result of a partial
gauge fixing; in fact, a gauge fixed version of Berkovits' theory resembling 
our approach has been explored by Iimori, Noumi, Okawa, and Torii 
\cite{Iimori,INOT}. Our regularization of the cubic vertex is inspired by their
work.

This paper is organized as follows. In section \ref{sec:WittenCub} we 
review Witten's superstring field theory up to cubic order and describe 
our regularization of the cubic vertex. In section \ref{sec:WittenQuar} we 
compute the quartic vertex by requiring that the BRST variation of the 
3-product cancel against the non-associativity of the 2-product. For this 
purpose it is useful to treat the picture changing operator $X$ as BRST
exact in the large Hilbert space. Then it is no longer guaranteed that the 
3-product will be independent of the $\xi$ zero mode. We determine a BRST 
exact correction which ensures that the 3-product is in the small Hilbert 
space. On the way, we find it useful to introduce some additional
multi-string products which play a central role in the recursion defining
higher vertices. In section \ref{sec:WittenQuin} we review some 
mathematical apparatus which relates multi-string products to coderivations 
on the tensor algebra, and use this language to streamline the computation 
of the quartic vertex and then the quintic vertex. In section 
\ref{sec:WittenAll} we derive a set of recursive equations which 
determine multi-string products to all orders. In section 
\ref{sec:amp} we show that the four-point amplitude derived from our 
theory agrees with the first quantized result. We end with some discussion.

\section{Witten's Theory up to Cubic order}
\label{sec:WittenCub}

The string field $\Psi$ is a Grassmann odd, ghost number $1$ and picture 
number $-1$ state in the boundary superconformal field theory of an open 
superstring quantized in a reference D-brane background. $\Psi$ is in the 
small Hilbert space, meaning it is independent of the zero mode of the 
$\xi$ ghost obtained upon bosonization of the $\beta\gamma$ system 
\cite{FMS}, or, equivalently, it is annihilated 
by the zero mode of the $\eta$ ghost, 
\begin{equation}\eta \Psi = 0,\end{equation}
where $\eta\equiv \eta_0$. The linear field equation is
\begin{equation}Q\Psi = 0,\end{equation}
where $Q\equiv Q_B$ is the worldsheet BRST operator. At picture $-1$, we can 
express on-shell states in Siegel gauge
\begin{equation}\Psi \sim ce^{-\phi}\mathcal{O}^\mathrm{m}(0),\end{equation}
where $\mathcal{O}^\mathrm{m}$ is a superconformal matter primary of 
dimension $1/2$. 

Let's explain a few sign conventions which are common in discussions of 
$A_\infty$ algebras, but are otherwise nonstandard in most discussions of 
open string field theory. Given a string field $A$ with Grassmann parity 
$\eps(A)$, we define its ``degree''
\begin{equation}\deg(A) \equiv 
\eps(A)+1\ \mathrm{mod}\ \mathbb{Z}_2.\end{equation}
The dynamical field $\Psi$ has even degree, though it corresponds to
a Grassmann odd vertex operator. We also define a $2$-product and 
symplectic form:
\begin{eqnarray}
m_2(A,B)\lineup \equiv (-1)^{\deg(A)}A*B,\label{eq:m2}\\
\omega(A,B)\lineup \equiv (-1)^{\deg(A)}\langle A,B\rangle.\label{eq:omega}
\end{eqnarray}
The 2-product is essentially the same as Witten's open string star product
except for the sign. Likewise, the symplectic form is essentially the
same as the BPZ inner product except for the sign. The main advantage of 
these sign conventions is that all multi-string products have the same 
(odd) degree as the BRST operator $Q$. In particular, $m_2$ adds one unit 
of degree when multiplying string fields:
\begin{equation}\deg(m_2(A,B)) = \deg(A)+\deg(B)+1.\end{equation}
These conventions slightly change the appearance of the familiar 
Chern-Simons axioms. The derivation property of $Q$ and the 
associativity of the star product take the form: 
\begin{eqnarray}
0\lineup =Q^2 A,\nonumber\\
0\lineup =Q m_2(A,B)+m_2(QA,B)+(-1)^{\deg(A)}m_2(A,QB),\nonumber\\
0\lineup =m_2(m_2(A,B),C)+(-1)^{\deg(A)}m_2(A,m_2(B,C)).
\end{eqnarray}
Rephrased in the appropriate language (to be described later), these 
relations can be understood as the statement that $Q$ and $m_2$ are 
nilpotent and anticommute. Finally, the symplectic form is 
BRST invariant
\begin{equation}0=\omega(QA,B)+ (-1)^{\deg(A)}\omega(A,Q B),\label{eq:omQ}
\end{equation}
and satisfies
\begin{equation}\omega(A,B) = -(-1)^{\deg(A)\deg(B)}\omega(B,A),\end{equation}
and so is (graded) antisymmetric.

Now let's discuss Witten's superstring field theory. Expanding the action up 
to cubic order gives\footnote{We normalize the ghost correlator 
$\langle c\d c\d^2c(x) e^{-2\phi}(y)\rangle=-2$ and set the open string 
coupling constant to one.}
\begin{equation}S=\frac{1}{2}\omega(\Psi,Q\Psi) 
+\frac{1}{3}\omega(\Psi,M_2(\Psi,\Psi))+...\ .\label{eq:action3}
\end{equation}
The 2-product $M_2$ above is different from the open string star product $m_2$.
In particular, the total picture must be $-2$ to obtain a nonvanishing 
correlator on the disk, so the 2-product $M_2(A,B)$ must have picture $+1$. 
The original proposal of Witten \cite{Witten} was to define $M_2$ using the 
open string star product with an insertion of the picture changing 
operator $X(z)=Q\cdot\xi(z)$ at the open string midpoint. Specifically, 
taking the sign inherited from \eq{m2},
\begin{equation}M_2(A,B)=X(i)m_2(A,B).\label{eq:Wittenprod}
\end{equation}
The problem is that repeated $M_2$-products are divergent due to a double 
pole in the $X$-$X$ OPE. This leads to a breakdown in 
gauge invariance and a divergence in the 4-point amplitude \cite{Wendt}. 
To avoid these problems we will make a more general ansatz:
\begin{equation}M_2(A,B) \equiv 
\frac{1}{3}\Big[X m_2(A,B)+m_2(XA,B) + m_2(A,XB)\Big],
\label{eq:M2}\end{equation}
where $X$ is a BPZ even charge of the picture changing operator:\footnote{We 
can choose $X$ to be BPZ even without loss of generality, since if we 
assume a cyclic vertex any BPZ odd component would cancel out.}
\begin{equation}X = \oint_{|z|=1} \frac{dz}{2\pi i}f(z) X(z).\end{equation}
The product $M_2$ now explicitly depends on a choice of 1-form $f(z)$, which 
describes how the picture changing is spread over the half-string 
overlaps of the Witten vertex. Provided $f(z)$ is holomorphic in some 
nondegenerate annulus around the unit circle, products of $X$ with itself 
are regular, and in particular the $4$-point amplitude is finite. Note that the
geometry of the cubic vertex \eq{M2} is the same as in Witten's open 
bosonic string field theory. This means that the propagator together 
with the cubic vertex already cover the bosonic moduli space of Riemann 
surfaces with boundary \cite{Zwcover}. Therefore higher vertices must be 
contact interactions without integration over bosonic moduli.

Since $X$ is BPZ even, the 1-form $f(z)$ satisfies
\begin{equation}f(z)=-\frac{1}{z^2}f\left(-\frac{1}{z}\right).\end{equation}
We also assume
\begin{equation}\oint_{|z|=1}\frac{dz}{2\pi i}f(z)=1,\label{eq:norm}
\end{equation}
since any other number could be absorbed into a redefinition of the open string
coupling constant. Perhaps the simplest choice of $X$ is the 
zero mode of the picture changing operator:
\begin{equation}X_0 =\oint_{|z|=1}\frac{dz}{2\pi i}\frac{1}{z}X(z).
\end{equation}
If we like, we can also choose $X$ so that it approaches Witten's 
singular midpoint insertion as a limit. For example we can take 
\begin{equation}f(z) = \frac{1}{z-i\lambda}
 - \frac{1}{z-\frac{i}{\lambda}},\label{eq:Witreg}
\end{equation}
which as $\lambda \to 1^-$ approaches a delta function localizing $X$ at
the midpoint. Note that the annulus of analyticity, 
\begin{equation}\lambda<|z|<\frac{1}{\lambda},\end{equation}
degenerates to zero thickness in the $\lambda\to 1^-$ limit. This is why
Witten's original vertex produces contact divergences. 

\section{Quartic Order}
\label{sec:WittenQuar}

The action constructed so far is not gauge invariant because 
the 2-product $M_2$ is not associative:
\begin{equation}M_2(M_2(A,B),C)+(-1)^{\deg(A)}M_2(A,M_2(B,C))\neq 0.
\end{equation}
To restore gauge invariance we search for a 3-product $M_3$, a 4-product
$M_4$, and so on so that the full set of multilinear maps satisfy the 
relations of an $A_\infty$ algebra. Using these multilinear maps to define 
higher vertices, the action 
\begin{equation}S=\frac{1}{2}\omega( \Psi,Q\Psi)+
\sum_{n=2}^\infty \frac{1}{n+1}\omega(
\Psi,M_n(\underbrace{\Psi,...,\Psi}_{n\ \mathrm{times}}))\end{equation}
is gauge invariant by construction. We offer a proof in appendix 
\ref{app:gauge}.

\begin{figure}
\begin{center}
\resizebox{4.2in}{1.6in}{\includegraphics{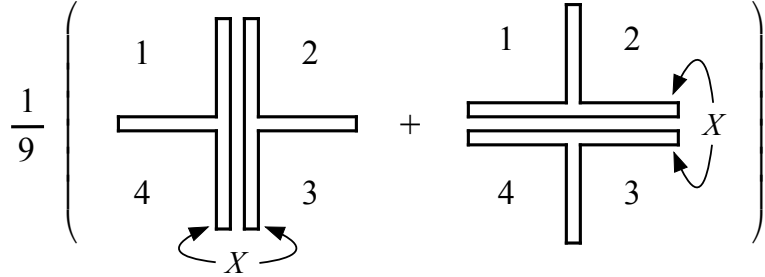}}
\end{center}
\caption{\label{fig:WittenSS1} Pictorial representation of the associator 
of $M_2$. We can take the numbers $1,2,3$ to represent the states which 
are multiplied, and $4$ to represent the output of the associator. The ``T'' 
shape represents a contour integral of $X$ surrounding the respective 
Witten vertex, and two factors of $\frac{1}{3}$ comes from the two vertices.}
\end{figure}

As a first step we construct the 3-product $M_3$ which defines the quartic
vertex. The first two $A_\infty$ relations say that $Q$ is nilpotent and a 
derivation of the 2-product $M_2$. The third relation characterizes the 
failure of $M_2$ to associate in terms of the BRST variation of $M_3$: 
\begin{eqnarray}0\lineup =M_2(M_2(A,B),C)+(-1)^{\deg(A)}M_2(A,M_2(B,C))+
QM_3(A,B,C)\nonumber\\ \lineup \ \ \ +M_3(QA,B,C)
+ (-1)^{\deg(A)}M_3(A,QB,C) 
+ (-1)^{\deg(A)+\deg(B)}M_3(A,B,QC).\nonumber\\
\label{eq:Ainf3}\end{eqnarray}
The last four terms represent the BRST variation of $M_3$ by
placing a $Q$ on each output of the quartic vertex. To visualize how to solve
for $M_3$, consider figure \ref{fig:WittenSS1}, which gives a schematic 
worldsheet picture the configuration of $X$ contour integrals in the $M_2$ 
associator. To pull a $Q$ off of the $X$ contours, it would clearly 
help if $X$ were a BRST exact quantity. In the large Hilbert space it is,
since we can write 
\begin{equation}X = [Q,\xi],\ \ \ \ \ \ \ \ 
\xi \equiv \oint_{|z|=1}\frac{dz}{2\pi i}f(z)\xi(z),
\end{equation}
where $\xi$ is the charge of the $\xi$-ghost defined by the 1-form $f(z)$. 
Now pulling a $Q$ out of the associator simply requires replacing one of the 
$X$ contours in each term with a $\xi$ contour. Since there are two $X$ 
contours in each term, there are two ways to do this, and by cyclicity we 
should sum both ways and divide by two.\footnote{We will say more about
 cyclicity in appendix \ref{app:cyclic}.} This is shown in figure 
\ref{fig:WittenSS2}. Translating this picture into an equation gives 
a solution for $M_3$: 
\begin{eqnarray}M_3(A,B,C) \lineup = \frac{1}{2}\Big[M_2(A,\Mb_2(B,C))
-(-1)^{\deg(A)}\Mb_2(A,M_2(B,C))\nonumber\\
\lineup \ \ \ +M_2(\Mb_2(A,B),C)-\Mb_2(M_2(A,B),C)\Big]+Q\text{-exact},
\label{eq:M3}\end{eqnarray}
where we leave open the possibility of adding a $Q$-exact piece (which would 
not contribute to the associator). $\Mb_2$ in this equation is a new object
that we call the {\it \dressed-2-product}:
\begin{equation}\Mb_2(A,B) \equiv \frac{1}{3}\Big[\xi m_2(A,B)-m_2(\xi A,B)
-(-1)^{\deg(A)}m_2(A,\xi B)\Big].\label{eq:Mb2}\end{equation}
This is essentially the same as $M_2$, only the $X$ contour has been replaced 
by a $\xi$ contour. The \dressed-2-product has even degree, and as required
its BRST variation is $M_2$:
\begin{equation}M_2(A,B)=Q\Mb_2(A,B)-\Mb_2(QA,B)-(-1)^{\deg(A)}\Mb_2(A,QB).
\label{eq:M2Q}\end{equation}
Acting $\eta$ on $\Mb_2$ gives yet another object which we call 
the {\it \bare-2-product}: 
\begin{equation} m_2(A,B)
=\eta\Mb_2(A,B)-\Mb_2(\eta A,B)-(-1)^{\deg{A}}\Mb_2(A,\eta B).\end{equation}
The \bare-2-product has odd degree. As it happens the \bare-2-product is the 
same as Witten's open string star product (with the sign of \eq{m2}). Both 
the \dressed-product and the \bare-product will have nontrivial higher-point 
generalizations.

\begin{figure}
\begin{center}
\resizebox{6.5in}{1.2in}{\includegraphics{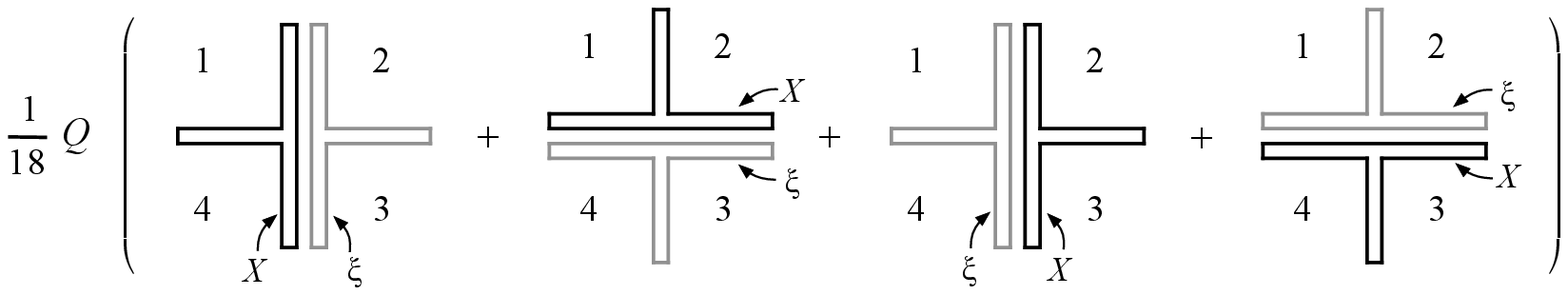}}
\end{center}
\caption{\label{fig:WittenSS2} Pictorial representation of the associator 
as a BRST exact quantity. The black ``T'' shape represents a contour integral
of $X$ around the Witten vertex and the grey ``T'' shape represents the 
corresponding contour integral of $\xi$. We have four terms since we 
require the quartic vertex to be cyclic.}
\end{figure}


So far the construction of the 3-product has seemed easy, essentially
because we have allowed ourselves to treat the 2-product as BRST exact.
But if the 2-product were ``truly'' BRST exact, then we would expect our theory
to produce a trivial $S$-matrix---in other words, it would be a complicated 
nonlinear rewriting of a free theory. A useful analogy to this situation 
is finding the first nonlinear correction to an infinitesimal gauge 
transformation. While this might be straightforward, usually constructing 
pure gauge solutions is not physically interesting. What makes our 
construction nontrivial is that the ``gauge transformation'' generating 
the cubic and quartic vertex lives in the large Hilbert space. And the result 
of the gauge transformation must be in the small Hilbert space. This suggests 
a structural analogy to solving the equations of motion of Berkovits 
superstring field theory. We will clarify the meaning of this analogy in 
appendix \ref{app:Linf}. 

\begin{figure}
\begin{center}
\resizebox{3.9in}{1.4in}{\includegraphics{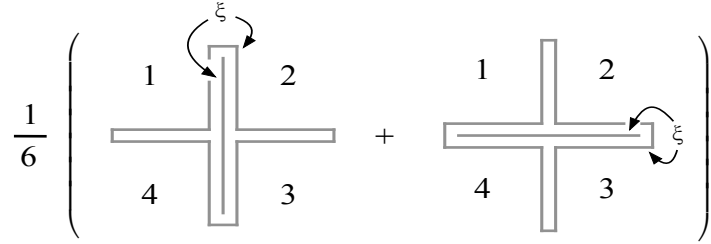}}
\end{center}
\caption{\label{fig:WittenSS5} Schematic picture of the $\xi$ contours
defining the \dressed-3-product. The vertical and horizontal lines inside 
the cross represents an insertion of $\xi$ between open string star products. 
The cross represents a sum of $\xi$ insertions acting on all external states.}
\end{figure}

This raises a central point: While we can introduce $\xi$ into our 
calculations as a formal convenience, consistency requires that all 
multilinear maps defining string vertices must be in the small 
Hilbert space. This is already true for $M_2$, but not yet true for $M_3$. 
For this reason we make use of our freedom to add a BRST exact piece in \eq{M3}
\begin{eqnarray}
Q\text{-exact}\lineup=\frac{1}{2}\Big[Q \Mb_3(A,B,C)- \Mb_3(QA,B,C) 
-(-1)^{\deg(A)}\Mb_3(A,QB,C)\nonumber\\
\lineup\ \ \ \ \ \ \ \ \ -(-1)^{\deg(A)+\deg(B)}
\Mb_3(A,B,QC)\Big],\label{eq:Qexact}
\end{eqnarray}
where $\Mb_3$ will be defined in such a way as to ensure that the total 
3-product is in the small Hilbert space. The object $\Mb_3$ will be called 
the {\it \dressed-3-product}. Now we require that $M_3$ is in the small 
Hilbert space:
\begin{eqnarray}0=\eta M_3(A,B,C) \lineup = 
\frac{1}{2}\Big[-(-1)^{\deg(A)}M_2(A,m_2(B,C))
-(-1)^{\deg(A)}m_2(A,M_2(B,C))\nonumber\\
\lineup \ \ \ -M_2(m_2(A,B),C)-m_2(M_2(A,B),C)\Big]
+\eta(Q\text{-exact}).
\end{eqnarray}
To avoid writing too many terms, we assume $A,B,C$ are in the small Hilbert 
space and the $Q$-exact piece is as in \eq{Qexact}. With some algebra
this simplifies to 
\begin{eqnarray}0=\eta M_3(A,B,C) \lineup= -\frac{1}{3}\Big[(-1)^{\deg(A)}
m_2(A,Xm_2(B,C))+m_2(Xm_2(A,B),C)\Big]\nonumber\\
\lineup\ \ \  +\eta(Q\text{-exact}). \end{eqnarray}
We now pull an overall $Q$ out of this equation. This replaces the $X$ 
insertion in the first two terms with a $\xi$ insertion:
\begin{eqnarray}\eta M_3(A,B,C) \lineup = Q\left(\frac{1}{3}\Big[
m_2(A, \xi m_2(B,C))+ m_2(\xi m_2(A,B),C)\Big]
-\frac{1}{2} \eta\Mb_3(A,B,C)\right)\nonumber\\
\lineup\ \ \ +\mathrm{other\ terms},\end{eqnarray}
where ``other terms'' take a similar form but with $Q$ acting on one of 
the three other external states. Since $\eta M_3$ should be zero, it is 
reasonable to assume that the \dressed-3-product $\Mb_3$ should satisfy 
\begin{eqnarray}\eta \Mb_3(A,B,C)\lineup = \frac{2}{3}\Big[m_2(A,\xi m_2(B,C))
+m_2(\xi m_2(A,B),C)\Big] \nonumber\\ 
\lineup \equiv m_3(A,B,C).
\label{eq:m3}\end{eqnarray}
The right hand side defines what we call the {\it \bare-3-product}, $m_3$. 
Of course, this equation is consistent only if the \bare-3-product happens to
be in the small Hilbert space. It is: Acting $\eta$ on $m_3$ gives the $m_2$ 
associator, which vanishes. Though equation \eq{m3} does not uniquely determine
$\Mb_3$, there is a natural solution: take $m_3$ and 
place a $\xi$ on each external state:
\begin{eqnarray}\Mb_3 \lineup \equiv \frac{1}{4}\Big[\xi m_3(A,B,C)
-m_3(\xi A,B,C)-(-1)^{\deg(A)}m_3(A,\xi B,C)\nonumber\\
\lineup\ \ \ \ \ \ -(-1)^{\deg(A)+\deg(B)}m_3(A,B,\xi C)\Big].
\label{eq:Mb3}\end{eqnarray}
Thus the \dressed-3-product is described by a configuration of 
$\xi$ contours shown in figure \ref{fig:WittenSS5}. This gives an explicit 
definition of the quartic vertex in the small Hilbert space 
consistent with gauge invariance.

\section{Quintic Order}
\label{sec:WittenQuin}

Performing all substitutions, the final expression for $M_3$ involves some 
30 terms with various combinations of $m_2$s, $X$s and $\xi$s acting on 
external states. At higher orders the vertices become even more complicated, 
and we need more economical notation. Therefore we explain a few conceptual 
and notational devices which are common in more mathematical discussions of 
$A_\infty$ algebras. See for example \cite{Kajiura} and references 
therein. Then we revisit the derivation of the quartic vertex, 
and continue on to the quintic vertex.

We are interested in multilinear maps taking $n$ copies of the BCFT state space
$\mathcal{H}$ into one copy. Such a map can be viewed as a linear operator 
from the $n$-fold tensor product of $\mathcal{H}$ into $\mathcal{H}$:
\begin{equation}b_n:\mathcal{H}^{\otimes n} \to \mathcal{H}.\end{equation}
Suppose we have a state in $\mathcal{H}^{\otimes n}$ of the form
\begin{equation}\Psi_1\otimes\Psi_2\otimes ... \otimes \Psi_n\in
\mathcal{H}^{\otimes n},\label{eq:tenPsi}\end{equation}
then $b_n$ acts on such a state as
\begin{equation}b_n(\Psi_1\otimes\Psi_2\otimes ... \otimes \Psi_n)
=b_n(\Psi_1,\Psi_2, ... ,\Psi_n),\label{eq:btenPsi}\end{equation}
where the right hand side is the multilinear map as denoted in previous 
sections. Since we can use the states \eq{tenPsi} to form a basis, 
\eq{btenPsi} defines the action of $b_n$ on the whole tensor 
product space.

Given $b_n$, define the following linear operator on 
$\mathcal{H}^{\otimes N\geq n}$:
\begin{equation}\mathbb{I}^{\otimes N-n-k}\otimes b_n\otimes
\mathbb{I}^{\otimes k}:\mathcal{H}^{\otimes N}\to \mathcal{H}^{\otimes N-n+1}.
\end{equation}
It acts on states of the form \eq{tenPsi} as 
\begin{eqnarray}\lineup
\mathbb{I}^{\otimes N-n-k}\otimes b_n\otimes
\mathbb{I}^{\otimes k}(\Psi_1\otimes\Psi_2\otimes ... \otimes \Psi_N)=
\nonumber\\
\lineup \ \ \ \ \ \  
(-1)^{\deg(b_n)(\deg(\Psi_1)+...+\deg(\Psi_{N-n-k}))}\times\nonumber\\
\lineup\ \ \ \ \  \ \ \ \ \ \ \ 
\Psi_1\otimes ...\otimes\Psi_{N-n-k} \otimes b_n(\Psi_{N-n-k+1},...,\Psi_{N-k})
\otimes\Psi_{N-k+1}\otimes...\otimes\Psi_N.
\end{eqnarray}
It acts in the obvious way: It leaves the tensor product of the first 
$N-n+k$ states untouched, multiplies the next $n$ states, and leaves the
tensor product of remaining $k$ states untouched. It also may produce a sign
from commuting $b_n$ past the first $N-n-k$ states.

With these ingredients we can define a natural action of $b_n$ or the tensor 
algebra:
\begin{equation}T\mathcal{H} = \mathcal{H}^{\otimes 0}\, \oplus\, 
\mathcal{H}\,\oplus\,\mathcal{H}^{\otimes 2}\,\oplus\,\mathcal{H}^{\otimes 3}
\,\oplus\, ...\ .\end{equation}
In this context we will denote the action of $b_n$ with a boldface $\b_n$:
\begin{equation}\b_n:T\mathcal{H}\to T\mathcal{H}.\end{equation}
$\b_n$ acts on the tensor algebra as a so-called 
{\it coderivation}.\footnote{To understand the origin of the term  
``coderivation,'' note that the tensor algebra $T\mathcal{H}$ has a 
natural ``coproduct''\begin{equation}\triangle:T\mathcal{H}\to T\mathcal{H}\otimes'T\mathcal{H}\end{equation} where we denote the tensor product symbol 
$\otimes'$ to distinguish it from the tensor product defining $T\mathcal{H}$. 
$\b_n$ is a coderivation in the sense that \begin{equation}\triangle\b_n 
= (\b_n\otimes'\mathbb{I}_{T\mathcal{H}}+\mathbb{I}_{T\mathcal{H}}
\otimes'\b_n)\triangle\end{equation} This is the ``dual'' of the Leibniz 
product rule. Though we borrow the terminology, we will not find a use for 
these extra structures. For further exposition, see \cite{Kajiura}.} 
We define $\b_n$ as follows: On the 
$\mathcal{H}^{\otimes N\geq n}$ component of the tensor algebra, we take
\begin{equation}\b_n \Psi
\equiv \sum_{k=0}^{N-n}\mathbb{I}^{\otimes N-n-k}\otimes b_n\otimes
\mathbb{I}^{\otimes k}\Psi,\ \ \ \ \Psi\in\mathcal{H}^{\otimes N\geq n}\subset
T\mathcal{H},\label{eq:coder}
\end{equation}
and on the $\mathcal{H}^{\otimes N<n}$ component, we take $\b_n$ to vanish.  
Naturally, on the $\mathcal{H}^{\otimes n}$ component, $\b_n=b_n$. So the 
coderivation $\b_n$ and multilinear map $b_n$ are isomorphic.

The advantage of this language is that it gives us a natural notion of 
``multiplication'' between multilinear maps. We just compose the 
corresponding coderivations. Particularly important are (graded) commutators 
of coderivations. With a little algebra, we can show that the commutator of 
two coderivations $\b_m,\b_n'$ derived from the maps
\begin{eqnarray}
\lineup b_m:\mathcal{H}^{\otimes m}\to\mathcal{H},\nonumber\\
\lineup b_n':\mathcal{H}^{\otimes n}\to\mathcal{H},
\end{eqnarray}
is a coderivation $[\b_m,\b_n']$ derived from the map\footnote{We always use
the bracket $[,]$ to denoted the commutator graded with respect to degree.}
\begin{equation}
[b_m,b_n']:\mathcal{H}^{m+n-1}\to\mathcal{H},
\end{equation}
with 
\begin{equation}[b_m,b_n']\equiv b_m \sum_{k=0}^{m-1}
\mathbb{I}^{\otimes m-1-k}\otimes b'_n\otimes\mathbb{I}^k-(-1)^{
\deg(b_m)\deg(b_m')}b_n'\sum_{k=0}^{n-1}
\mathbb{I}^{\otimes n-1-k}\otimes b_m\otimes\mathbb{I}^k.\label{eq:cocom}
\end{equation}
The sums in this equation are closely related to the multitude of terms which 
appear in formulas for the 3-product. This notation allows us to keep 
track of these terms in a very economical fashion.

Consider for example the first three $A_\infty$ relations:
\begin{eqnarray}
0\lineup =Q^2 A,\\
0\lineup =QM_2(A,B)+M_2(QA,B)+(-1)^{\deg(A)}M_2(A,QB),\\
0\lineup =M_2(M_2(A,B),C)+(-1)^{\deg(A)}M_2(A,M_2(B,C))+
QM_3(A,B,C)\nonumber\\ \lineup \ \ \ +M_3(QA,B,C)
+ (-1)^{\deg(A)}M_3(A,QB,C) 
+ (-1)^{\deg(A)+\deg(B)}M_3(A,B,QC).\nonumber\\
\end{eqnarray}
Recalling \eq{btenPsi}, we can ``factor out'' the string fields $A,B,C$:
\begin{eqnarray}
0\lineup =Q^2, \\
0\lineup =QM_2+M_2(Q\otimes\mathbb{I}+\mathbb{I}\otimes Q),\\
0\lineup =M_2(M_2\otimes\mathbb{I}+\mathbb{I}\otimes M_2)
+QM_3+M_3(Q\otimes \mathbb{I}\otimes\mathbb{I}
+\mathbb{I}\otimes Q\otimes \mathbb{I} + \mathbb{I}\otimes\mathbb{I}\otimes Q),
\nonumber\\
\end{eqnarray}
Now from \eq{cocom} we recognize these terms as commutators of 
coderivations. The first three $A_\infty$ relations reduce to
\begin{eqnarray}
0\lineup =\frac{1}{2}[\Q,\Q],\\
0\lineup =[\Q,\M_2],\\
0\lineup =[\Q,\M_3]+\frac{1}{2}[\M_2,\M_2].\label{eq:2Ainf3}
\end{eqnarray}

Now let's return to the quartic vertex. The 2-product and \dressed-2-product
are defined
\begin{eqnarray}
M_2 \lineup \equiv 
\frac{1}{3}\Big(Xm_2+m_2(X\otimes\mathbb{I}+\mathbb{I}\otimes X)\Big),\\
\Mb_2\lineup \equiv \frac{1}{3}\Big(\xi m_2-m_2(\xi\otimes\mathbb{I}
+\mathbb{I}\otimes \xi)\Big),
\end{eqnarray}
and satisfy
\begin{eqnarray}
\M_2 = [\Q,\Mbb_2],\\
\m_2 = [\n,\Mbb_2],
\end{eqnarray}
where $\m_2$ is the \bare-2-product. Following \eq{M3}, the 3-product is 
expressed
\begin{equation}\M_3 = \frac{1}{2}\Big([\Q,\Mbb_3]+[\M_2,\Mbb_2]\Big).
\label{eq:2M3}\end{equation}
where $\Mbb_3$ is the \dressed-3-product. Now its easy to plug into 
\eq{2Ainf3} and check the relevant $A_\infty$ relation. Taking the 
commutator with $\Q$ the first term in \eq{2M3} drops out since $[\Q,\Q]=0$. 
Using the Jacobi identity the second term gives $-\frac{1}{2}[\M_2,\M_2]$, 
which cancels against the $\frac{1}{2}[\M_2,\M_2]$ term in \eq{2Ainf3}.

Now we need to make sure $\M_3$ is in the small Hilbert space. Acting with 
$\n$ we find
\begin{eqnarray}0=[\n,\M_3]\lineup 
= \frac{1}{2}\Big(-[\Q,[\n,\Mbb_3]]-[\M_2,\m_2]\Big),\nonumber\\
\lineup 
= \frac{1}{2}\Big(-[\Q,[\n,\Mbb_3]]-[[\Q,\Mbb_2],\m_2]\Big),\nonumber\\
\lineup 
= \frac{1}{2}\big[\Q,-[\n,\Mbb_3]+[\m_2,\Mbb_2]\big].
\end{eqnarray}
Since this should vanish, we assume 
\begin{equation}[\n, \Mbb_3]=\m_3\equiv [\m_2,\Mbb_2], \label{eq:2m3}
\end{equation}
where $\m_3$ is the \bare-3-product. This is consistent since 
$\m_3$ is in the small Hilbert space:
\begin{equation}[\n,\m_3] = -[\m_2,\m_2]=0,\end{equation}
where we used associativity of $\m_2$. Thus we can define the 
\dressed-3-product by placing a $\xi$ on each output of the \bare-3-product:
\begin{equation}\Mb_3 = \frac{1}{4}\Big(\xi m_3-m_3(
\xi\otimes\mathbb{I}\otimes\mathbb{I}+\mathbb{I}\otimes\xi\otimes\mathbb{I}
+\mathbb{I}\otimes\mathbb{I}\otimes\xi)\Big).\label{eq:2Mb3}\end{equation}
Via \eq{2M3}, this completely determines the four vertex. 

Now we claim that a similar procedure extends to higher orders. Just to see 
it work in the next example, let's construct the quintic vertex. The 
relevant $A_\infty$ relation is
\begin{equation}0=[\Q,\M_4]+[\M_2,\M_3].\end{equation}
The solution is
\begin{equation}\M_4=\frac{1}{3}\Big([\Q,\Mbb_4]+[\M_2,\Mbb_3]+[\M_3,\Mbb_2]
\Big),
\end{equation}
where $\Mbb_4$ is the {\it \dressed-4-product}. To check, compute:
\begin{eqnarray}[\Q,\M_4] \lineup = 
\frac{1}{3}\Big(-[\M_2,[\Q,\Mbb_3]]+[[\Q,\M_3],\Mbb_2]-[\M_3,\M_2]\Big),
\nonumber\\
\lineup = \frac{1}{3}\Big(-\big[\M_2,2\M_3 -[\M_2,\Mbb_2]\big]+
\big[-\half[\M_2,\M_2],\Mbb_2\big]
-[\M_3,\M_2]\Big),\nonumber\\
\lineup = \frac{1}{3}\Big(-2[\M_2,\M_3]+[\M_2,[\M_2,\Mbb_2]]
-[\M_2,[\M_2,\Mbb_2]]
-[\M_2,\M_2]\Big),\nonumber\\
\lineup = \frac{1}{3}\Big(-3[\M_2,\M_3]\Big),\nonumber\\
\lineup = -[\M_2,\M_3].
\end{eqnarray}
In the first step we used the Jacobi identity and $[\Q,\Q]=0$, $[\Q,\M_2]=0$ 
and $[\Q,\Mbb_2]=\M_2$. In the second step we used the $A_\infty$ relation for 
$\M_3$ and used \eq{2M3} to solve for $[\Q,\Mbb_3]$. The remaining steps use
the Jacobi identity. Since we want $\M_4$ to be in the small Hilbert space 
we demand
\begin{eqnarray}
0= [\n,\M_4]\lineup = 
\frac{1}{3}\Big(-[\Q,[\n,\Mbb_4]]-[\M_2,\m_3]-[\M_3,\m_2]\Big),\nonumber\\
\lineup =\frac{1}{3}\left(-[\Q,[\n,\Mbb_4]]-[\M_2,\m_3]
-\frac{1}{2}\big[[\Q,\Mbb_3]+[\M_2,\Mbb_2],\m_2\big]\right),\nonumber\\
\lineup =\frac{1}{3}\left(-[\Q,[\n,\Mbb_4]]-[\M_2,\m_3]
-\frac{1}{2}[[\M_2,\m_2],\Mbb_2]+\frac{1}{2}[\M_2,[\m_2,\Mbb_2]]
+\frac{1}{2}[\Q,[\m_2,\Mbb_3]]\right),\nonumber\\
\lineup =\frac{1}{3}\left(-[\Q,[\n,\Mbb_4]]-\frac{1}{2}[\m_3,\M_2]
+\frac{1}{2}[[\Q,\m_3],\Mbb_2]
+\frac{1}{2}[\Q,[\m_2,\Mbb_3]]\right),\nonumber\\
\lineup =\frac{1}{3}\left[\Q,\left(-[\n,\Mbb_4]+\frac{1}{2}[\m_3,\Mbb_2]
+\frac{1}{2}[\m_2,\Mbb_3]\right)\right].
\end{eqnarray}
In the second equation we substituted \eq{2M3} in place of $\M_3$. In the third
we used the Jacobi identity. In the fourth we substituted the definition of 
$\m_3$, and in the fifth we pulled out a $\Q$. Since this should vanish, we 
assume
\begin{equation}[\n,\Mbb_4] = \m_4 \equiv \frac{1}{2}\Big([\m_3,\Mbb_2]
+[\m_2,\Mbb_3]\Big),\end{equation}
where $\m_4$ is the {\it \bare-4-product}. Consistently, $\m_4$ is in the 
small Hilbert space:
\begin{eqnarray}[\n,\m_4]\lineup = [\m_3,\m_2],\nonumber\\
\lineup = [[\m_2,\Mbb_2],\m_2],\nonumber\\
\lineup = [[\m_2,\m_2],\Mbb_2] -[[\m_2,\Mbb_2],\m_2],\nonumber\\
\lineup = [[\m_2,\m_2],\Mbb_2]- [\n,\m_4].\nonumber\\
\lineup = 0\end{eqnarray}
Therefore the \dressed-4-product can be constructed by placing a $\xi$ on 
each output of $m_4$:
\begin{equation}\Mb_4 = \frac{1}{5}\Big(\xi m_4 - m_4(
\xi\otimes\mathbb{I}\otimes\mathbb{I}\otimes\mathbb{I}+\mathbb{I}\otimes \xi
\otimes\mathbb{I}\otimes\mathbb{I}+\mathbb{I}\otimes\mathbb{I}\otimes \xi
\otimes\mathbb{I}+\mathbb{I}\otimes\mathbb{I}\otimes\mathbb{I}\otimes\xi)
\Big).\end{equation}
This completely fixes the theory up to quintic order.

\section{Witten's Theory to All Orders}
\label{sec:WittenAll}

Now we are ready to discuss the construction of vertices to all orders.
The $n$-th $A_\infty$ relation reads
\begin{equation}
0=[\M_n,\M_1]+[\M_{n-1},\M_2]+...+[\M_2,\M_{n-1}]+[\M_1,\M_n],
\end{equation}
where $\M_1\equiv \Q$. To express all such relations in a compact form, it is 
useful to introduce a generating function $\M(t)$:
\begin{equation}\M(t) \equiv \sum_{n=0}^\infty t^n \M_{n+1},\label{eq:Mt}
\end{equation}
where $t$ is some parameter. Then the full set of $A_\infty$ relations is 
equivalent to the equation
\begin{equation}[\M(t),\M(t)]=0.\end{equation}
The $n$th relation is found by expanding this equation in a power series and
reading off the coefficient of $t^{n-1}$.

The solution we're after takes the form
\begin{equation}\M_{n+2} = \frac{1}{n+1}\sum_{k=0}^{n} [\M_{n-k+1},\Mbb_{k+2}].
\label{eq:Mnp1}\end{equation}
If we know the products up to $\M_{n+1}$, and the \dressed-products up to 
$\Mbb_{n+2}$, this equation determines the next product $\M_{n+2}$. The proof 
is as follows. Define a generating function for the \dressed-products:
\begin{equation}\Mbb(t) = \sum_{n=0}^\infty t^n \Mbb_{n+2}\end{equation}
Then the recursive formula \eq{Mnp1} follows from the $t^n$ component
of the differential equation
\begin{equation}\frac{d}{dt}\M(t) = [\M(t),\Mbb(t)].\label{eq:Mdif}
\end{equation}
This equation implies
\begin{equation}\frac{d}{dt}[\M(t),\M(t)] = 2[[\M(t),\M(t)],\Mbb(t)].
\label{eq:AinfMb}\end{equation}
Let
\begin{equation}[\M(t),\M(t)]_{n+1}=\sum_{k=0}^n[\M_{n-k+1},\M_{k+1}],
\end{equation}
be the combination of $\M$s appearing in the $n+1$st $A_\infty$ relation, or 
equivalently the coefficient of $t^n$ in the power series expansion of 
$[\M(t),\M(t)]$. Then equation \eq{AinfMb} implies a recursive formula for 
these coefficients:
\begin{equation}[\M(t),\M(t)]_{n+2} = \frac{2}{n+1}\sum_{k=0}^{n}
[[\M(t),\M(t)]_{n-k+1},\Mbb_{k+2}].\end{equation}
If $[\M(t),\M(t)]_k$ vanishes for $1\leq k\leq n+1$, then this formula implies 
that it must vanish for $k=n+2$. So all we have to do is show that 
$[\M(t),\M(t)]_k$ vanishes for $k=1$. It does because
\begin{equation}[\M(t),\M(t)]_1=[\Q,\Q]=0.\end{equation}
This completes the proof that \eq{Mnp1} implies the $A_\infty$ relations.

Next consider the \bare-products $\m_n$. For the moment we will ignore 
the possible identification between $\m_n$ and $[\n,\Mbb_n]$. Rather, we will
define the \bare-products in terms of the recursive formula
\begin{equation}\m_{n+3} = \frac{1}{n+1}\sum_{k=0}^{n}[\m_{n-k+2},\Mbb_{k+2}].
\label{eq:mnp2}\end{equation}
If we know the \bare-products up to $\m_{n+2}$ and the \dressed-products up to
$\Mbb_{n+2}$, this determines the next \bare-product $\m_{n+3}$. We can check 
that this formula matches our previous calculation of the \bare-3-product and 
\bare-4-product. Suppose that we define a generating function for the 
\bare-products
\begin{equation}\m(t) = \sum_{n=0}^\infty t^n\m_{n+2}.\end{equation}
Then \eq{mnp2} implies the differential equation
\begin{equation}\frac{d}{dt}\m(t)=[\m(t),\Mbb(t)].\end{equation}
Using a similar argument as just given below \eq{AinfMb}, we can prove  
\begin{eqnarray}[\m(t),\m(t)]\lineup =0,\\
\ [\m(t),\M(t)]\lineup = 0
\end{eqnarray}
recursively from the identities $[\m_2,\m_2]=0$ and $[\m_2,\Q]=0$. In 
components of $t^n$,
\begin{eqnarray}\sum_{k=0}^n[\m_{n-k+2},\m_{k+2}]\lineup =0,\label{eq:Ainfm}\\
\sum_{k=0}^n [\m_{n-k+2},\M_{k+1}]\lineup = 0.
\end{eqnarray}
This means that the products and \bare-products form a pair of mutually 
commuting $A_\infty$ algebras.

This much is true regardless of our choice of \dressed-products $\Mbb_k$. 
What fixes $\Mbb_k$ is the additional condition
\begin{equation}[\n,\Mbb_{k+2}]=\m_{k+2}.\label{eq:etMb}\end{equation}
We construct a solution to this condition recursively as follows. First 
note that $[\eta,\Mbb_2]=\m_2$ by definition. Second, suppose that we have 
constructed a solution to \eq{etMb} up to $\m_{n+2}$ and $\Mbb_{n+2}$. 
Then it follows that the \bare-product $\m_{n+3}$ is in the small Hilbert 
space:
\begin{equation}[\n,\m_{n+3}] 
= -\frac{1}{n+1}\sum_{k=0}^{n}[\m_{n-k+2},\m_{k+2}]=0,\end{equation}
where we used the recursive equation \eq{mnp2} and the $A_\infty$ relations
\eq{Ainfm}. Now define the $n+3$rd \dressed-product:
\begin{equation}\Mb_{n+3} \equiv \frac{1}{n+4}\left(\xi m_{n+3}-m_{n+3}
\sum_{k=0}^{n+2}\mathbb{I}^{\otimes n+2-k}\otimes \xi\otimes
\mathbb{I}^{\otimes k}\right).\label{eq:Mb}\end{equation}
Since $\m_{n+3}$ is in the small Hilbert space, this implies 
\begin{equation}\ [\n,\Mbb_{n+3}]=\m_{n+3}.\end{equation}
Proceeding this way inductively, we find a solution to \eq{etMb} for all $k$.

Next we have to show how this construction implies that all products defining 
vertices are in the small Hilbert space. Acting $\n$ on the differential 
equation \eq{Mdif} for $\M$ gives
\begin{eqnarray}\frac{d}{dt}[\n,\M(t)] \lineup = 
[[\n,\M(t)],\Mbb(t)]-[\M(t),\m(t)],\nonumber\\
\lineup = [[\n,\M(t)],\Mbb(t)],
\end{eqnarray}
where we used \eq{etMb} and the fact that the $A_\infty$ algebras of 
$\M$ and $\m$ commute. The $t^n$ component of this differential equation 
implies the recursive formula
\begin{equation}[\n,\M_{n+2}] = \frac{1}{n+1}\sum_{k=0}^{n} 
[[\n,\M_{n-k+1}],\Mbb_{k+2}].\end{equation}
Note that $\M_1=\Q$ commutes with $\eta$. And this equation implies that if 
all of the products up to $\M_{n+1}$ are in the small Hilbert space, the 
next product $\M_{n+2}$ is also in the small Hilbert space. Thus we have a 
complete solution of the $A_\infty$ relations defining Witten's superstring 
field theory.

\begin{figure}
\begin{center}
\resizebox{4in}{1.9in}{\includegraphics{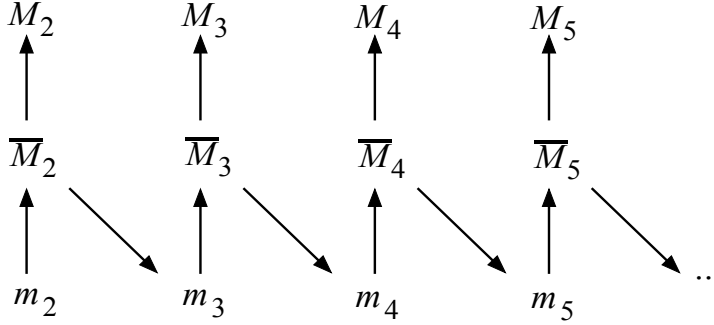}}
\end{center}
\caption{\label{fig:WittenSS3} General pattern of recursion defining all 
higher products. At any stage, we always start with the \bare-product and 
proceed to derive the \dressed-product. Next, we can either find the 
``true'' product that defines the vertex, or proceed to the next 
\bare-product and start the process over.}
\end{figure}

The construction we have provided is recursive. Suppose we have determined 
all products, \bare-products, and \dressed-products up to $\M_n,\m_n$ and 
$\Mbb_n$. To proceed to the next order, first we construct the $n+1$st 
\bare-product $\m_{n+1}$ from equation \eq{mnp2}. Next we construct the 
$n+1$st \dressed-product $\Mbb_{n+1}$ from equation \eq{Mb}. Finally, using
$\Mbb_{n+1}$ we construct the $n+1$st product $\M_{n+1}$ via \eq{Mnp1}, 
or we can proceed to the next order and compute the $n+2$nd \bare-product 
$\m_{n+2}$, starting the process over. The general pattern of recursion 
is illustrated in figure \ref{fig:WittenSS3}.

Our solution to the $A_\infty$ relations depends on the following assumptions:
\begin{description}\itemindent0pt
\item{(1)} $Q$ and $\eta$ are nilpotent and anticommute.
\item{(2)} $Q$ and $\eta$ are derivations of the product $m_2$.
\item{(3)} $\eta$ has a homotopy $\xi$ satisfying $[\eta,\xi]=1$.
\item{(4)} $m_2$ is associative.
\end{description}
Within the context of these assumptions we can construct a slightly more 
general solution by adding an $\eta$ closed piece to $\xi$. This can have 
the effect of replacing $X$ in the cubic vertex with a slightly more general 
operator. Aside from this, perhaps the most interesting assumption to drop 
is associativity of $m_2$. This might be useful, for example, 
for constructing a theory based on a cubic vertex with worldsheet strips 
attached to each output, as is done in open-closed bosonic string field 
theory \cite{ZwOpCl}.

The solution of the $A_\infty$ relations is not unique. The non-uniqueness 
can be characterized by our freedom to add an $\eta$ closed piece 
to $\Mbb_n$ at each order. Perhaps the most nontrivial aspect of our 
construction is that despite this non-uniqueness we were able to find a 
natural definition of each vertex, without having to make additional choices 
at each order. In other words, we found a way to ``fix the gauge.''

\section{Four-point Amplitudes}
\label{sec:amp}

It is interesting to see how our regularization of Witten's theory reproduces 
the familiar first-quantized scattering amplitudes. Here we focus 
on the generic four-point amplitude. The general case can probably be 
treated in a similar fashion.\footnote{Similar computations of 
four-point amplitudes in gauge-fixed Berkovits superstring field theory 
appear in \cite{INOT}.}

We start with the color-ordered 4-point amplitude expressed
in the form:
\begin{equation}
A_4^{\1st}(\Psi_1,\Psi_2,\Psi_3,\Psi_4) = -\int_0^1 dt\Big\langle
\Big(X_0\cdot \Psi_1(0)\Big)\Big(b_{-1}X_0\cdot\Psi_2(t)\Big)\Psi_3(1)\,
\Psi_4(\infty)
\Big\rangle_{UHP}.
\end{equation}
Here $\Psi_1,..,\Psi_4$ are on-shell vertex operators in the $-1$ picture,
and the correlator is evaluated in the small Hilbert space on the upper half
plane. We denote the amplitude with the superscript ``1st'' to indicate that 
this is the first quantized amplitude, not (yet) the string field theory 
result. As far as bosonic moduli are concerned, this amplitude is 
structurally the same as in the bosonic string, and following 
\cite{Giddings} we can reexpress it using the open string star product and 
the Siegel gauge propagator in the $s$- and $t$-channels:
\begin{eqnarray}
\lineup \!\!\!\!\!\!\!\!\!\!A_4^{\1st}(\Psi_1,\Psi_2,\Psi_3,\Psi_4)
=\nonumber\\
\lineup\!\!\!\!\!-\omega\left( X_0\Psi_1, m_2\left(X_0\Psi_2,
\frac{b_0}{L_0}m_2(\Psi_3,\Psi_4)\right)\right)-
\omega\left(X_0\Psi_1,m_2\left(\frac{b_0}{L_0}m_2(X_0\Psi_2,\Psi_3),\Psi_4
\right)\right).\nonumber\\
\label{eq:1stamp0}\end{eqnarray}
This is the form of the amplitude we want to compare with Witten's superstring
field theory.

Now consider the 4-point amplitude derived from the Lagrangian:
\begin{eqnarray}
\lineup\!\!\!\!\!\!\!\!\!\!\!\!\! A_4(\Psi_1,\Psi_2,\Psi_3,\Psi_4)=\nonumber\\
\lineup \!\!\!-\omega\left(\Psi_1, M_2\left(\Psi_2,
\frac{b_0}{L_0}M_2(\Psi_3,\Psi_4)\right)\right)-
\omega\left(\Psi_1,
M_2\left(\frac{b_0}{L_0}M_2(\Psi_2,\Psi_3),\Psi_4\right)\right)\nonumber\\
\lineup\!\!\!+\omega\Big(\Psi_1,M_3(\Psi_2,\Psi_3,\Psi_4)\Big).
\end{eqnarray}
The amplitude can be viewed as a multilinear map from the four-fold
tensor product of the physical state space into complex numbers
\begin{equation}\langle A_4|: \mathcal{H}_Q^{\otimes 4}\to \mathbb{C},
\end{equation}
where $\mathcal{H}_Q\subset\mathcal{H}$ is the subspace of states annihilated 
by $Q$. Pulling $\Psi_1,...,\Psi_4$ off to the right we can then express the
amplitude
\begin{eqnarray}\langle A_4| = \langle \omega|\left(\mathbb{I}\otimes 
M_2\left(-\mathbb{I}\otimes \frac{b_0}{L_0}M_2-\frac{b_0}{L_0}M_2\otimes
\mathbb{I}\right)+\mathbb{I}\otimes M_3\right),\end{eqnarray}
where $\langle \omega|:\mathcal{H}^{\otimes 2}\to\mathbb{C}$ is the symplectic
form. We can write this using the coderivations derived from $M_2$ and $M_3$: 
\begin{equation}\langle A_4| = \langle \omega|\mathbb{I}\otimes \pi_1
\left(-\M_2\frac{b_0}{L_0}\M_2 +\M_3\right),\label{eq:2ndamp}\end{equation}
where we use $\frac{b_0}{L_0}\M_2$ to denote the coderivation derived from 
the map $\frac{b_0}{L_0}M_2$. The symbol $\pi_1$ means we let the 
coderivations act on the last three states, and select the component of 
the output in $\mathcal{H}$. We can also write the first quantized amplitude 
\eq{1stamp0}
\begin{equation}\langle A_4^\1st| = -\langle \omega|\mathbb{I}\otimes
\pi_1\left(\m_2\frac{b_0}{L_0}\m_2\right)(X_0\otimes X_0\otimes 
\mathbb{I}\otimes\mathbb{I}).\label{eq:1stamp}
\end{equation}
Let's prove that BRST exact states decouple. Suppose the 
first state $\Psi_1$ is BRST exact. Pulling the $Q$ off $\Psi_1$ and acting on 
$\langle A_4|$ gives
\begin{eqnarray}
\langle A_4|Q\otimes \mathbb{I}\otimes\mathbb{I}\otimes\mathbb{I}
\lineup =-\langle A_4|Q\otimes\pi_1
\left(-\M_2\frac{b_0}{L_0}\M_2 +\M_3\right),\nonumber\\
\lineup = \langle \omega|\mathbb{I}\otimes\pi_1\left(
-\Q\M_2\frac{b_0}{L_0}\M_2+\Q\M_3\right),
\end{eqnarray}
where we used the fact that $Q$ is BPZ odd: $\langle\omega|\mathbb{I}\otimes 
Q = -\langle\omega|Q\otimes\mathbb{I}$. Since the other three states are 
BRST closed, we can write the second factor as a commutator with $\Q$:
\begin{eqnarray}
\langle A_4|Q\otimes \mathbb{I}\otimes\mathbb{I}\otimes\mathbb{I}
\lineup = \langle \omega|\mathbb{I}\otimes\pi_1\left(\left[\Q,
-\M_2\frac{b_0}{L_0}\M_2+\M_3\right]\right),\nonumber\\
\lineup = \langle \omega|\mathbb{I}\otimes\pi_1\big(\M_2\M_2+[\Q,\M_3]\big),
\nonumber\\
\lineup = \langle \omega|\mathbb{I}\otimes\pi_1\left(\frac{1}{2}[\M_2,\M_2]
+[\Q,\M_3]\right),\nonumber\\
\lineup = 0.
\end{eqnarray}
This vanishes as a result of the $A_\infty$ relation for $M_2$ and $M_3$. 
Similarly, BRST exact states decouple from the first quantized 
amplitude \eq{1stamp} because of associativity of $m_2$.

Now we want to show that the field theory amplitude \eq{2ndamp} and the 
first-quantized amplitude \eq{1stamp} are identical. For this purpose it is 
helpful to pass to the large Hilbert space, since this allows us to analyze
individual terms which appear in the 3-product $M_3$ separately. Let us 
denote the large Hilbert space $\mathcal{H}_L$, and the subspace of 
$\eta$-closed states $\mathcal{H}_\eta\subset\mathcal{H}_L$. There is an 
obvious isomorphism between the small Hilbert space $\mathcal{H}$ and 
$\mathcal{H}_\eta$:
\begin{equation}L:\mathcal{H}\to\mathcal{H}_\eta.\end{equation}
We take the states on either side to be defined by the same vertex 
operator. However, the symplectic form on $\mathcal{H}$ and $\mathcal{H}_\eta$
are different; the later requires saturation by the $\xi$ zero mode. For our
calculation, it is useful to define the symplectic form on the small Hilbert
space $\omega$ in terms of the symplectic form on the large Hilbert space 
$\omega_L$ as follows:\footnote{This identification assumes that the 
basic ghost correlator in the large Hilbert space is normalized 
$\langle \xi c\d c\d^2c e^{-2\phi}\rangle = 2$. Note that the sign is 
opposite from our chosen normalization of the basic correlator in the small 
Hilbert space.}
\begin{equation}\langle \omega| 
= \langle \omega_L|(\mathbb{I}\otimes \xi)(L\otimes L).
\label{eq:om}\end{equation}
If $b_n$ is a multilinear map which commutes with $\eta$, this implies 
the relation
\begin{equation}\langle\omega|\mathbb{I}\otimes b_n=(-1)^{\deg(b_n)}
\langle\omega_L|
(\mathbb{I}\otimes b_n)
(\mathbb{I}^{\otimes k}\otimes \xi\otimes\mathbb{I}^{n-k})L^{\otimes n+1},
\label{eq:ximove}\end{equation}
so we can place $\xi$ on any input of the multilinear map as needed.

Passing to the large Hilbert space, the amplitude now acts on the 4-fold 
tensor product of BRST invariant states in $\mathcal{H}_\eta$, which we 
denote $\mathcal{H}_{Q\eta}$:
\begin{equation}\langle A_{4,L}|:\mathcal{H}_{Q\eta}^{\otimes 4}\to\mathbb{C},
\ \ \ \ \mathcal{H}_{Q\eta}\subset\mathcal{H}_\eta\subset\mathcal{H}_L.
\end{equation}
Taking care of the $\xi$ zero mode, the field theory amplitude \eq{2ndamp} 
now takes the form
\begin{equation}\langle A_{4,L}| = \langle \omega_L|\mathbb{I}\otimes 
\xi\pi_1\left(
-\M_2\frac{b_0}{L_0}\M_2+\M_3\right),
\end{equation}
where we used \eq{om}. Since we are in the large Hilbert space, we are 
free to use our definition of the vertices in terms of dressed and bare 
products. Write $\M_2 = [\Q,\Mbb_2]$ in the first term and pull $[\Q,\cdot]$ 
past the propagator:
\begin{eqnarray}\langle A_{4,L}|\lineup = 
\langle \omega_L|\mathbb{I}\otimes \xi\pi_1\left(
-\frac{1}{2}\left[\Q,\Mbb_2\frac{b_0}{L_0}\M_2\right]-
\frac{1}{2}\left[\Q,\M_2\frac{b_0}{L_0}\Mbb_2\right]
-\frac{1}{2}[\M_2,\Mbb_2]+\M_3\right),
\nonumber\\
\lineup = \langle \omega_L|\mathbb{I}\otimes X \pi_1\left(
-\frac{1}{2}\Mbb_2\frac{b_0}{L_0}\M_2+
\frac{1}{2}\M_2\frac{b_0}{L_0}\Mbb_2\right)
+\langle \omega_L|\mathbb{I}\otimes
\xi\pi_1\left(-\frac{1}{2}[\M_2,\Mbb_2]+\M_3\right).\nonumber\\
\end{eqnarray}
In the second step we moved the $\Q$ commutator past the $\xi$ insertion 
to act on external states. Note that $-\frac{1}{2}[\M_2,\Mbb_2]$ already 
cancels one term in $\M_3$. In the first pair of terms above $\xi$ only 
appears in the dressed 
2-product $\Mbb_2$. Using \eq{ximove} we can move the $\xi$s out of 
$\Mbb_2$ onto the second entry of the symplectic form. This leaves the 
bare 2-product $\m_2$:
\begin{equation}
\langle A_{4,L}|= \langle \omega_L|\mathbb{I}\otimes X \xi \pi_1\left(
-\frac{1}{2}\m_2\frac{b_0}{L_0}\M_2-
\frac{1}{2}\M_2\frac{b_0}{L_0}\m_2\right)
+\langle \omega_L|\mathbb{I}\otimes
\xi\pi_1\left(-\frac{1}{2}[\M_2,\Mbb_2]+\M_3\right).
\end{equation} 
Now we repeat this process a second time; Write $\M_2=[\Q,\Mbb_2]$ and 
pull $[\Q,\cdot]$ past the propagator:
\begin{eqnarray}
\langle A_{4,L}| \lineup = \langle \omega_L|\mathbb{I}\otimes X \xi \pi_1\left(
\frac{1}{2}\left[\Q,\m_2\frac{b_0}{L_0}\Mbb_2\right]-
\frac{1}{2}\left[\Q,\Mbb_2\frac{b_0}{L_0}\m_2\right]
-\frac{1}{2}[\m_2,\Mbb_2]\right)
\nonumber\\
\lineup\ \ \ \ \ +\langle \omega_L|\mathbb{I}\otimes
\xi\pi_1\left(-\frac{1}{2}[\M_2,\Mbb_2]+\M_3\right).
\end{eqnarray} 
We pick up a term $[\m_2,\Mbb_2]$, which happens to be the 
bare-3-product $\m_3$. Moving $Q$ past the 
$\xi$ insertion gives
\begin{eqnarray}
\langle A_{4,L}| \lineup = \langle \omega_L|\mathbb{I}\otimes X^2 \pi_1\left(
-\frac{1}{2}\m_2\frac{b_0}{L_0}\Mbb_2-
\frac{1}{2}\Mbb_2\frac{b_0}{L_0}\m_2\right)\nonumber\\
\lineup\ \ \ -\langle \omega_L|\mathbb{I}\otimes X\xi \pi_1 \left(\frac{1}{2}
\m_3\right)
+\langle \omega_L|\mathbb{I}\otimes
\xi\pi_1\left(-\frac{1}{2}[\M_2,\Mbb_2]+\M_3\right).
\end{eqnarray} 
In the first term, use \eq{ximove} to move the $\xi$ out of $\Mbb_2$ onto 
the second input of $\omega_L$. In the second term, use \eq{ximove} to move
the $\xi$ from the second input of $\omega_L$ back into the bare-3-product 
$\m_3$, turning it into the dressed 3-product $\Mbb_3$:
\begin{eqnarray}
\langle A_{4,L}| \lineup = \langle \omega_L|\mathbb{I}\otimes X^2 \xi
\pi_1\left(-\m_2\frac{b_0}{L_0}\m_2\right)
 -\langle \omega_L|\mathbb{I}\otimes X\pi_1 \left(\frac{1}{2}\Mbb_3\right)
\nonumber\\
\lineup\ \ \ \ \ +\langle \omega_L|\mathbb{I}\otimes
\xi\pi_1\left(-\frac{1}{2}[\M_2,\Mbb_2]+\M_3\right),\nonumber\\
\lineup = \langle \omega_L|\mathbb{I}\otimes X^2 \xi
\pi_1\left(-\m_2\frac{b_0}{L_0}\m_2\right) +\langle \omega_L|\mathbb{I}\otimes
\xi\pi_1\left(-\frac{1}{2}[\Q,\Mbb_3]-\frac{1}{2}[\M_2,\Mbb_2]+\M_3\right).
\nonumber\\
\end{eqnarray} 
The last three terms cancel by the definition of $\M_3$. Moving back 
to the small Hilbert space, we have therefore shown
\begin{equation}\langle A_4| = -\langle \omega|X^2\otimes
\pi_1\left(\m_2\frac{b_0}{L_0}\m_2\right).\end{equation}
This is almost the first quantized amplitude, except $X$ may be different 
from the zero mode $X_0$, and it acts twice on the first input rather than once
on the first and once on the second input. But the difference between $X$ and 
$X_0$ is a BRST exact, and the change moving $X_0$ to the second output
is also BRST exact. Since external states are on shell and $m_2$ is 
associative, these changes do not effect the amplitude. Therefore
\begin{equation}\langle A_4|= 
-\langle \omega|\mathbb{I}\otimes
\pi_1\left(\m_2\frac{b_0}{L_0}\m_2\right)(X_0\otimes X_0\otimes 
\mathbb{I}\otimes\mathbb{I})=\langle A_4^\1st|.\end{equation}
and the string field theory 4-point amplitude agrees with the first quantized
result.

\section{Discussion}

We have succeeded in constructing an explicit and nonsingular
covariant superstring field theory in the small Hilbert space. Virtually by 
construction, the action satisfies the classical BV master equation,
\begin{equation}\{S,S\}=0,\end{equation}
once we relax the ghost number constraint on the string field. To quantize 
the theory, we need to incorporate the Ramond sector. 
There are a couple of different approaches we could take to this problem. 
One suggested by Berkovits \cite{BerkRamond} is to distribute the degrees 
of freedom of the Ramond string field between picture $-\frac{1}{2}$ and 
picture $-\frac{3}{2}$, which necessarily breaks manifest covariance. 
One might also try to regulate Witten's original kinetic term for the 
Ramond string field, which has a midpoint insertion of the inverse picture 
changing operator $Y$. Then we would 
have to see how this extra operator could be incorporated into the 
$A_\infty$ structure. Once the Ramond sector is included, we would be 
in good shape to understand the role of closed strings in quantum open 
string field theory.

Another variation we can consider is adding stubs to the cubic vertex. 
Then the higher vertices would necessarily require integration over bosonic 
moduli. It would be interesting to understand the interplay between the 
picture changing insertions and the $A_\infty$ structure related to 
integration over bosonic moduli. Once this is understood it is plausible 
that closed Type II superstring field theory could be constructed in a 
similar manner. Previous formal attempts to construct such a theory have 
been stymied by the lack of a well-posed minimal area problem on supermoduli 
space \cite{Korbinian}. A recent construction of Type II closed superstring 
field theory in the large Hilbert space may also provide input on this problem 
\cite{Matsunaga}.

Our construction is purely algebraic. We have not analyzed how the 
vertices and propagators cover the supermoduli space of the disk with NS 
boundary punctures. Understanding this would undoubtedly provide insight 
into the foundations of superstring field theory. 

Considering that our theory is formulated in the small Hilbert space, the 
large Hilbert space plays a surprisingly prominent role. This strongly 
suggests a relation to Berkovits' open superstring field theory. It would 
be interesting if our formulation could be derived by gauge fixing 
the Berkovits theory \cite{Iimori,INOT}. For one thing, there has been recent 
notable progress in understanding classical solutions in the Berkovits 
theory \cite{Erler}, and it would be pleasing to incorporate these results 
in a unified formalism.

\bigskip

\noindent {\bf Acknowledgments}

\bigskip

\noindent Many thanks to M. Kroyter for organizing the conference SFT 2012
in Jerusalem which triggered our interest in this problem. I.S. specifically
acknowledges interesting discussions with B. Zwiebach at that conference. 
T.E. would like to thank Y. Okawa for explaining ideas which provided the 
basis for our regularization of the cubic vertex. T.E. also thanks C. 
Maccaferri for comments on the second draft of this paper. This project 
was supported in parts by the DFG Transregional Collaborative Research Centre 
TRR 33, the DFG cluster of excellence Origin and Structure of the Universe as 
well as the DAAD project 54446342.

\begin{appendix}

\section{Gauge Invariance}
\label{app:gauge}

We would like to explain why the $A_\infty$ relations imply gauge 
invariance of the action. Of course, gauge invariance follows from having a 
solution to the BV master equation, and often having a solution to the BV 
master equation is of greater interest. But it is nice to see a direct 
proof of gauge invariance without invoking Batalin-Vilkovisky machinery.

The classical action is
\begin{equation}S=\sum_{n=0}^\infty \frac{1}{n+2}\omega(
\Psi,M_{n+1}(\underbrace{\Psi,...,\Psi}_{n+1\ \mathrm{times}})),
\end{equation}
and the infinitesimal gauge transformation is
\begin{equation}\delta\Psi = \sum_{n=0}^\infty\sum_{k=0}^n 
M_{n+1}(\underbrace{\Psi,...,\Psi}_{n-k\ \mathrm{times}},\Lambda,
\underbrace{\Psi,...,\Psi}_{k\ \mathrm{times}}),\end{equation}
where $\Lambda$ is the gauge parameter. To prove gauge invariance we must 
assume that the vertices are cyclic:
\begin{eqnarray}
\omega(M_{n+1}(\Psi_1,...,\Psi_{n+1}),\Psi_{n+2})\lineup =
(-1)^{\deg(\Psi_1)(\deg(\Psi_2)+...
+\deg(\Psi_{n+2}))}\nonumber\\
\lineup\ \ \ \ \ \ \ \ \ \ \times
\omega(M_{n+1}(\Psi_2,...,\Psi_{n+2}),\Psi_1).\label{eq:cyclic}\end{eqnarray}
Products that satisfy this condition are said to define a {\it cyclic
$A_\infty$ algebra} \cite{Kajiura}. We will demonstrate that our products are 
cyclic in appendix \ref{app:cyclic}. Since the vertices are cyclic, when 
we vary the action we can bring all of the $\delta\Psi$s to the first 
entry of the symplectic form, producing a factor of $n+2$. Thus
\begin{equation}\delta S=\sum_{n=0}^\infty \omega(
\delta\Psi,M_{n+1}(\underbrace{\Psi,...,\Psi}_{n+1\ \mathrm{times}})).
\end{equation}
Plugging in $\delta\Psi$
\begin{equation}\delta S=\sum_{m,n=0}^\infty \sum_{l=0}^m\omega(
M_{m+1}(\underbrace{\Psi,...,\Psi}_{m-l\ \mathrm{times}},\Lambda,
\underbrace{\Psi,...,\Psi}_{l\ \mathrm{times}}),
M_{n+1}(\underbrace{\Psi,...,\Psi}_{n+1\ \mathrm{times}})).
\end{equation}
Now use cyclicity to get the $\Lambda$ to the second entry of $\omega$:
\begin{equation}
\delta S =-\sum_{m,n=0}^\infty \sum_{l=0}^m\omega(
M_{m+1}(\underbrace{\Psi,...,\Psi}_{l\ \mathrm{times}},M_{n+1}(\underbrace{\Psi,...,\Psi}_{n+1\ \mathrm{times}}),
\underbrace{\Psi,...,\Psi}_{m-l\ \mathrm{times}}),\Lambda).\end{equation}
With a little notational rearrangement,
\begin{equation}\delta S= 
-\sum_{m,n=0}^\infty\omega\left(M_{m+1}\left(\sum_{l=0}^m
\mathbb{I}^{\otimes l}\otimes M_{n+1}\otimes\mathbb{I}^{\otimes m-l}\right)
\Psi^{\otimes m+n+1},\Lambda \right).
\end{equation}
Relabeling the sums,
\begin{equation}\delta S= 
-\sum_{N=0}^\infty\omega\left(\sum_{k=0}^N M_{N-k+1}\left(\sum_{l=0}^{N-k}
\mathbb{I}^{\otimes l}\otimes M_{k+1}\otimes\mathbb{I}^{\otimes N-k-l}\right)
\Psi^{\otimes N+1},\Lambda \right).
\end{equation}
The $A_\infty$ relations imply
\begin{equation}
\sum_{k=0}^N M_{N-k+1}\left(\sum_{l=0}^{N-k}
\mathbb{I}^{\otimes l}\otimes M_{k+1}\otimes\mathbb{I}^{\otimes N-k-l}\right)
=0.
\end{equation}
This is simply a reexpression of the $A_\infty$ relations for 
coderivations acting on the $\mathcal{H}^{\otimes N+1}$ subspace of the tensor
algebra:
\begin{equation}\sum_{k=0}^N [ \M_{N-k+1},\M_{k+1}] = 0.\end{equation}
Therefore
\begin{equation}\delta S =0,\end{equation}
and the action is gauge invariant.

\section{Cyclicity of Vertices}
\label{app:cyclic}

Though we have shown that our vertices satisfy the $A_\infty$ relations, 
we did not prove cyclicity. Cyclicity of the vertex in the form \eq{cyclic}
follows from antisymmetry of the symplectic form together with the relation
\begin{equation}
\omega(M_n(\Psi_1,...,\Psi_{n}),\Psi_{n+1}) =
-(-1)^{\deg(\Psi_1)}
\omega(\Psi_1,M_{n}(\Psi_2,...,\Psi_{n+1})).\label{eq:exchange}
\end{equation}
Striping off the string fields, we can express this equation in the form
\begin{equation}\langle \omega | \mathbb{I}\otimes M_n = 
-\langle \omega|M_n\otimes \mathbb{I}.\end{equation}
where $\langle \omega|:\mathcal{H}^{\otimes 2}\to\mathbb{C}$ is the 
symplectic form. In this sense, the multi-string products should be BPZ 
odd, like the BRST operator. From the nature of our construction of the 
products, cyclicity can be inferred from two facts:

\begin{description}
\item{{\bf Fact 1:}} Let $b_m$ and $b_n'$ be two BPZ odd multilinear maps. 
Then the commutator
\begin{equation}[b_m,b_n']\equiv b_m \sum_{k=0}^{m-1}
\mathbb{I}^{\otimes m-1-k}\otimes b'_n\otimes\mathbb{I}^k-(-1)^{
\deg(b_m)\deg(b_m')}b_n'\sum_{k=0}^{n-1}
\mathbb{I}^{\otimes n-1-k}\otimes b_m\otimes\mathbb{I}^k
\end{equation}
is a BPZ odd multilinear map. 
\item{{\bf Fact 2:}} Let $b_m$ be a BPZ odd multilinear map and $C$ a BPZ 
even operator. Then the ``anticommutator'' defined
\begin{equation}\{C,b_m\}\equiv C b_m +(-1)^{\deg(C)\deg(b_m)}
b_m\left(\sum_{k=0}^{m-1}\mathbb{I}^{\otimes m-k-1}\otimes
C\otimes\mathbb{I}^{\otimes k}\right)\label{eq:anticom}\end{equation}
is a BPZ odd multilinear map.
\end{description}

\noindent We put ``anticommutator'' in quotes since
the anticommutator of $\b_m$ and $\bf{C}$ is not a coderivation. Note that
fact 2 applies specifically when $C$ is a BPZ even operator, and does not 
generalize to BPZ even multilinear maps.

\begin{proof} Let's start with fact 1. Plugging in \eq{cocom} we find the 
expression
\begin{eqnarray}\langle \omega|\mathbb{I}\otimes [b_m,b_n']\lineup 
=\sum_{k=0}^{m-1}\langle\omega|\mathbb{I}\otimes b_m (\mathbb{I}^{\otimes m-1-k}\otimes b'_n\otimes\mathbb{I}^k)\nonumber\\
\lineup\ \ \ \ \ \ -(-1)^{\deg(b_m)\deg(b_m')}\sum_{k=0}^{n-1}
\langle\omega|\mathbb{I}\otimes b_n'(\mathbb{I}^{\otimes n-1-k}
\otimes b_m\otimes\mathbb{I}^k).\nonumber\\\end{eqnarray}
Now we want to pull the $b$s onto the first input of $\omega$:
\begin{eqnarray}
\langle \omega|\mathbb{I}\otimes [b_m,b_n']\lineup
=-\sum_{k=0}^{m-2}\langle\omega|b_m(\mathbb{I}^{\otimes m-1-k}\otimes b'_n\otimes\mathbb{I}^{k})\otimes \mathbb{I} - \langle\omega|b_m\otimes b_n'
\nonumber\\
\lineup\ \ \  +(-1)^{\deg(b_m)\deg(b_m')}\left(\sum_{k=0}^{n-2}
\langle\omega|b_n'(\mathbb{I}^{\otimes n-1-k}\otimes b_m\otimes\mathbb{I}^k)
\otimes\mathbb{I}+
\langle\omega|b_n'\otimes b_m\right).\nonumber\\
\label{eq:step2}
\end{eqnarray}
Now we have two extra terms with $b$s acting on both inputs of $\omega$.
Again we have to pull a $b$ onto the first input:
\begin{eqnarray}\lineup- \langle\omega|b_m\otimes b_n'
+(-1)^{\deg(b_m)\deg(b_m')}\langle\omega|b_n'\otimes b_m = \nonumber\\
\lineup\ \ \ \ \ \ \ \ \ \ \ \ \ \ \ \ \ \ \ \ \ \ 
(-1)^{\deg(b_m)\deg(b_m')}\langle\omega|b_n'(b_m\otimes\mathbb{I}^{\otimes n-1})\otimes \mathbb{I} - \langle\omega|b_m(b_n'\otimes\mathbb{I}^{\otimes m-1})
\otimes \mathbb{I}.\nonumber\\
\end{eqnarray}
This fills a missing entry in the sums in \eq{step2}. So we find 
\begin{eqnarray}
\langle \omega|\mathbb{I}\otimes [b_m,b_n']\lineup
=-\sum_{k=0}^{m-1}\langle\omega|b_m(\mathbb{I}^{\otimes m-1-k}\otimes b'_n\otimes\mathbb{I}^{k})\otimes \mathbb{I}
\nonumber\\
\lineup\ \ \  \ \ \ \ +(-1)^{\deg(b_m)\deg(b_m')}\sum_{k=0}^{n-1}
\langle\omega|b_n'(\mathbb{I}^{\otimes n-1-k}\otimes b_m\otimes\mathbb{I}^k)
\otimes\mathbb{I},\nonumber\\
\lineup =-\langle\omega|[b_m,b_n'] \otimes\mathbb{I},
\end{eqnarray}
which establishes fact 1. Now for fact 2. Plugging in,
\begin{equation}\langle\omega|\mathbb{I}\otimes\{C,b_m\}= 
\langle\omega|\mathbb{I}\otimes C b_m +(-1)^{\deg(C)\deg(b_m)}
\sum_{k=0}^{m-1}\langle\omega|\mathbb{I}\otimes b_m(\mathbb{I}^{\otimes n-k-1}
\otimes C\otimes\mathbb{I}^{\otimes k}).\end{equation}
In the first term we pull the $C$ and then the $b$ onto the first input, 
and in the second term we pull the $b$ onto the first input:
\begin{eqnarray}
\langle\omega|\mathbb{I}\otimes\{C,b_m\}\lineup = 
-(-1)^{\deg(C)\deg{b_m}}\langle\omega|b_m(C\otimes\mathbb{I}^{\otimes m-1})
\otimes \mathbb{I} \nonumber\\
\lineup \ \ \ -(-1)^{\deg(C)\deg(b_m)}\left(
\sum_{k=0}^{m-2}\langle\omega|b_m(\mathbb{I}^{\otimes m-k-1}
\otimes C\otimes\mathbb{I}^{\otimes k})\otimes \mathbb{I}
+\langle\omega|b_m\otimes C\right).\nonumber\\
\end{eqnarray}
The first term fills a missing entry in the sum in the second term, and in the
third term we pull the $C$ onto the second input. Thus
\begin{eqnarray}\langle\omega|\mathbb{I}\otimes\{C,b_m\}\lineup = 
-\langle\omega|C b_m \otimes \mathbb{I}
 -(-1)^{\deg(C)\deg(b_m)}\sum_{k=0}^{m-1}\langle\omega|
b_m(\mathbb{I}^{\otimes m-k-1}\otimes C\otimes\mathbb{I}^{\otimes k})
\otimes \mathbb{I},\nonumber\\
\lineup = -\langle\omega|\{C,b_m\}\otimes\mathbb{I},
\end{eqnarray}
which establishes fact 2.
\end{proof}

\noindent All of our higher products are constructed from previous ones 
using operations covered by facts 1 and 2. Then since $Q$ and $m_2$ define
cyclic vertices, all vertices are cyclic.

\section{$L_\infty$ gauge transformations}
\label{app:Linf}

Earlier we mentioned that our vertices are derived as a kind of 
``gauge transformation'' of the free theory through the large Hilbert 
space. This is analogous to how Berkovits' superstring field theory 
derives solutions to the Chern-Simons equations of motion as a ``gauge 
transformation'' in the large Hilbert space. This is an interesting 
point, and deserves some explanation.

Suppose we have a set of multilinear maps $b_1, b_2, b_3,...$ acting on some 
graded vector space satisfying the relations of an $A_\infty$ algebra. We 
can add their coderivations to form
\begin{equation}\b = \b_1+\b_2+\b_3+\b_4+...\ .\end{equation}
The coderivation $\b$ incorporates all of the multilinear maps into a single 
entity. If we want to recover the map $b_n$, we simply act $\b$ on 
$\mathcal{H}^{\otimes n}$ and look at the component of the output in 
$\mathcal{H}$. The $A_\infty$ relations can be expressed in a compact form
\begin{equation}[\b,\b]=0.\end{equation}
Thus the whole $A_\infty$ algebra can be described by a single nilpotent 
coderivation $\b$ on the tensor algebra.

We are interested in deformations of this $A_\infty$ structure. Thus we look 
for a new coderivation $\b' = \b + \c$ which is nilpotent.
This implies that the perturbation $\c$ must satisfy the Maurer-Cartan equation
\begin{equation}d_\b \c + \frac{1}{2}[\c,\c] = 0,\label{eq:coEOM}\end{equation}
where 
\begin{equation}d_\b \equiv [\b,\cdot]\end{equation}
is called the Hochschild differential. Noting $\frac{1}{2}[\c,\c]=\c^2$, 
this looks just like the Chern-Simons equations of motion. There is a subtle 
difference however; coderivations do not naturally form an associative algebra,
since the composition of two coderivations is not generally a coderivation.
Rather, coderivations form a Lie algebra, and in particular, together with 
the Hochschild differential, a differential graded Lie algebra---the simplest 
example of an $L_\infty$ algebra. Therefore equation \eq{coEOM} is actually 
more closely analogous to the equations of motion of closed string field 
theory. 

Equation \eq{coEOM} has many solutions, some of which are 
``gauge equivalent.'' Gauge equivalence in this context is implemented by a 
so-called {\it $L_\infty$ gauge transformation}. It takes the form
\begin{equation}\c' = \g^{-1}(d_\b + \c)\g,\end{equation}
where $\g$ is an element of the group formally obtained by exponentiating 
coderivations of even degree. The solutions of the Maurer-Cartan equation 
\eq{coEOM}, modulo $L_\infty$ gauge transformations, defines the moduli space 
of $A_\infty$ structures around $\b$. If $\b$ describes the multilinear maps 
of open bosonic string field theory, then the moduli space formally represents 
the set of consistent closed string backgrounds 
\cite{MuensterSachs}.\footnote{Since gauge invariance requires cyclic 
vertices, we should be careful to consider only perturbations which define 
cyclic $A_\infty$ algebras.} This is somewhat 
subtle, since finite deformations of the closed string background usually 
change the nature of the boundary conformal field theory, and it is not clear
in what sense the deformed $A_\infty$ structure acts on the same tensor 
algebra. However, if $\b$ corresponds to Witten's open bosonic string field 
theory, it has been shown that solutions of the linearized equation,
\begin{equation}d_\b \c=0,\end{equation}
precisely reproduce the closed string cohomology \cite{MoellerSachs}. 
Therefore the Maurer-Cartan equation can see consistent closed string 
backgrounds at least in an infinitesimal neighborhood of the reference 
bulk conformal field theory. 

Now consider a 1-parameter family of $A_\infty$ algebras:
\begin{equation}\b(t) = \b+\c(t),\ \ \ \c(0)=0.\end{equation}
The Maurer-Cartan equation implies that infinitesimal variation 
$\epsilon\frac{d}{dt}\b(t)$ along the trajectory should be annihilated 
by the Hochschild differential at time $t$:
\begin{equation}d_{\b(t)}\frac{d}{dt}\b(t)=0.\end{equation}
Furthermore, if the solutions $\b(t)$ are gauge equivalent, then the 
variation along the trajectory should be trivial in the Hochschild cohomology: 
\begin{equation}\frac{d}{dt}\b(t) = d_{\b(t)}(\mathrm{something}).
\end{equation}
Now we are ready to explain the sense in which our vertices are derived as a 
gauge transformation from the free theory. Taking the
products $Q,M_2,M_3,...$ we can build the coderivation 
\begin{equation}\M = \M_1+\M_2+\M_3+\M_4+...\ .\end{equation}
This expression is the same as the generating function \eq{Mt},
\begin{equation}\M(t) = \sum_{n=0}^\infty t^n\M_{n+1},\end{equation}
evaluated at $t=1$. And at $t=0$, $\M(t)$ reduces to 
\begin{equation}\M(0)=\Q.\end{equation}
Thus the generating function $\M(t)$ defines a 1-parameter family of 
$A_\infty$ algebras connecting the free theory to Witten's superstring 
field theory with coupling constant set to $1$. $\M(t)$ satisfies the 
differential equation \eq{Mdif}, which can be written in the form:
\begin{equation}\frac{d}{dt}\M(t) = d_{\M(t)}\Mbb(t).\end{equation}
But this is exactly the statement that infinitesimal variations 
along the trajectory are trivial in the Hochschild cohomology. Therefore, 
we have constructed Witten's superstring field theory, described by $\M$, as 
a finite $L_\infty$ gauge transformation of the free theory, described by $\Q$.
The \dressed-products $\Mbb(t)$ are the infinitesimal gauge parameters which 
generate the trajectory connecting these two theories. Explicitly, 
the finite gauge transformation takes the form
\begin{equation}\M = \Q + \g^{-1}d_\Q \g,\end{equation}
where
\begin{equation}\g = \mathcal{P}\exp\left[\int_0^1 dt\,\Mbb(t)\right],
\end{equation}
and $\mathcal{P}$ denotes the path ordered exponential.

Of course, $\Mbb(t)$ does not generate a ``true'' gauge transformation since
it is in the large Hilbert space. And the theories $\M(t)$ differ by having a 
factor of $t^n$ in front of $M_{n+1}$, which can be viewed as adjusting the 
coupling constant---a feature of the closed string background which cannot
be changed by an $L_\infty$ gauge transformation. The thing that makes this
work is that the $L_\infty$ gauge transformation is in the large Hilbert space,
while our theory is defined in the small Hilbert space. Specifically, we have 
solved the equation
\begin{equation}[\n,\g^{-1}d_\Q \g]=0.\label{eq:BerkEOM}\end{equation}
Structurally, this is identical to the equations of motion in Berkovits' open 
superstring field theory. Only the solutions of \eq{BerkEOM} represent 
consistent open superstring field theories, rather than open string 
backgrounds. 

It is interesting that our solution of the $A_\infty$ relations proceeds more
naturally through the analogue of the Berkovits equations of motion 
\eq{BerkEOM}, rather than the Maurer-Cartan equation \eq{coEOM}. This is 
opposite to what happens in most analytic studies of classical solutions in
Berkovits' string field theory. Usually it is more natural to start with 
the solution of the Chern-Simons equations of motion (which are similar to 
those of the bosonic string) and then lift to a solution of the Berkovits
theory 
\cite{supermarg,Okawa_real_super_marg,FK,KO_super,FKeq,simple_super_marg}. 
Therefore our solution of the $A_\infty$ relations gives a possibly 
useful technique for constructing new classical solutions in Berkovits' 
superstring field theory.

\end{appendix}

\end{document}